\documentclass[twocolumn,amsmath,amssymb]{revtex4}
\usepackage{epsfig}

\begin{document}


\title{ New Concept for Testing General Relativity:\\[2pt]
The Laser Astrometric Test of Relativity (LATOR) Mission}
 
\author{{Slava G. Turyshev},\footnote{Electronic address:
        turyshev@jpl.nasa.gov}$^a$  
{Michael Shao},\footnote{Electronic address: 
        mshao@huey.jpl.nasa.gov}$^a$  
        and 
{Kenneth Nordtvedt, Jr.}\footnote{Electronic address: kennordtvedt@imt.net}$^b$ \\[2pt]
~~~~ }

\affiliation{$^a$Jet Propulsion Laboratory, California Institute
of  Technology, Pasadena, CA 91109 U.S.A.\\
$^b$Northwest Analysis, 118 Sourdough Ridge Road, Bozeman, MT 59715 U.S.A.}



\begin{abstract} 

This paper discusses a Fundamental physics experiment that will test relativistic gravity at the accuracy better than the effects of the second order in the gravitational field strength, $\propto G^2$. The Laser Astrometric Test Of Relativity (LATOR) mission uses laser interferometry between two micro-spacecraft whose lines of sight pass close by the Sun to accurately measure deflection of light in the solar gravity.  The key element of the experimental design is a redundant geometry optical truss provided by a long-baseline (100 m) multi-channel stellar optical interferometer placed on the International Space Station (ISS). The spatial interferometer is used for measuring the angles between the two spacecraft and for orbit determination purposes. In Euclidean geometry, determination of a triangle's three sides determines any angle therein; with gravity changing the optical lengths of sides passing close by the Sun and deflecting the light, the Euclidean relationships are overthrown. The geometric redundancy enables LATOR to measure the departure from Euclidean geometry caused by the solar gravity field to a very high accuracy.
LATOR will not only improve the value of the parameterized post-Newtonian (PPN) $\gamma$ to unprecedented levels of accuracy of 1 part in 10$^{8}$, it will also reach ability to measure effects of the next post-Newtonian order ($c^{-4}$) of light deflection resulting from gravity's intrinsic non-linearity.  The solar quadrupole moment parameter, $J_2$, will be measured with high precision, as well as a variety of other relativistic effects including Lense-Thirring precession.  LATOR will lead to very robust advances in the tests of Fundamental physics: this mission could discover a violation or extension of general relativity, or reveal the presence of an additional long range interaction in the physical law.  There are no analogs to the LATOR experiment; it is unique and is a natural culmination of solar system gravity experiments.

\end{abstract}    


\maketitle

\section{INTRODUCTION}

Einstein's general theory of relativity (GR) began with its empirical success in 1915 by explaining the anomalous perihelion precession of Mercury's orbit, using no adjustable theoretical parameters.  Shortly thereafter,  Eddington's 1919 observations of star lines-of-sight during a solar eclipse confirmed the doubling of the deflection angles predicted by GR as compared to Newtonian and Equivalence Principle arguments.  From these beginnings, the general theory of relativity has been verified at ever higher accuracy. Thus, microwave ranging to the Viking Lander on Mars yielded accuracy  $\sim$0.1\% in the tests of GR \cite{[47],[54],[55]}.  The astrometric observations of quasars on the solar background performed with Very-Long Baseline Interferometry (VLBI) improved the accuracy of the tests of gravity to $\sim$ 0.03\% \cite{[23],[48],[32]}.  Lunar Laser Ranging (LLR),  a continuing legacy of the Apollo program, provided $\sim$ 0.01\% verification of the general relativity via precision measurements of the lunar orbit \cite{[39],[41],[45],[46],[63],[12n],Nordvedt,[62]}. Finally, the recent experiments with the Cassini spacecraft resulted a remarkable test of relativistic gravity accurate to $\sim$ 0.003\% \cite{cassini_ber}. As a result, by now not only the ``non-relativistic,'' Newtonian regime is well understood, but also the first ``post-Newtonian'' approximation is also well-studied, making general relativity the standard theory of gravity when astrometry and spacecraft navigation are concerned. 

However, the continued inability to merge gravity with quantum mechanics, and recent observations in cosmology indicate that the pure tensor gravity of general relativity needs modification or augmentation.  Recent work in scalar-tensor extensions of gravity which are consistent with present cosmological models \cite{[12a],[12]} motivate new searches for very small deviations of relativistic gravity in the solar system, at levels of 10$^{-5}$ to 10$^{-7}$ of the post-Newtonian effects or essentially to achieve accuracy that enables measurement of the effects of the 2nd order in the gravitational field strength ($\propto G^2$).  This will require a several order of magnitude improvement in experimental precision from present tests. At the same time, it is well understood that the ability to measure the second order light deflection term  would enable one to demonstrate even higher accuracy in measuring the first order effect, which is of the utmost importance for the gravitational theory and is the challenge for the 21st century Fundamental physics.

When the light deflection in solar gravity is concerned, the magnitude of the first order effect as predicted by GR for the light ray just grazing the limb of the Sun is $\sim1.75$ arcsecond. The effect varies inversely with the impact parameter. The second order term is almost six orders of magnitude smaller resulting in  $\sim 3.5$ microarcseconds ($\mu$as) light deflection effect, and which falls off inversely as the square of the light ray's impact parameter \cite{[22],[26],[49],[50],[40]}.  The gravitomagnetic frame-dragging term (effect in which both the orientation and trajectory of objects in orbit around a body are altered by the gravity of the body's rotation; it was studied by Lense and Thirring in 1918) is $\pm 0.7 ~\mu$as, and contribution of the solar quadrupole moment, $J_2$, is sized as 0.2 $\mu$as (using the value of the solar quadrupole moment $J_2\simeq10^{-7}$). The small magnitudes of the effects emphasize the fact that, among the four forces of nature, gravitation is the weakest interaction; it acts at very long distances and controls the large-scale structure of the universe, thus, making the precision tests of gravity a very challenging task. 

The LATOR  mission concept will directly address the challenges discussed above. The test will be performed in the solar gravity field using optical interferometry between two micro-spacecraft.  Precise measurements of the angular position of the spacecraft will be made using a fiber coupled multi-chanel led optical interferometer on the International Space Station (ISS) with a 100 m baseline. The primary objective of the LATOR Mission will be to measure the gravitational deflection of light by the solar gravity to accuracy of 0.1 picoradians, which corresponds to $\sim$10 picometers on a 100 m interferometric baseline. 

In conjunction with laser ranging among the spacecraft and the ISS, LATOR will allow measurements of the gravitational deflection by a factor of 30,000 better than has previously been accomplished. In particular, this mission will not only measure the key parameterized post-Newtonian (PPN) $\gamma$ to unprecedented levels of accuracy of one part in 10$^8$, it will also reach ability to measure the next post-Newtonian order ($c^{-4}$) of light deflection resulting from gravity's intrinsic non-linearity.  As a result, this experiment will measure values of other PPN parameters such as $\delta$ to 1 part in $10^3$ (never measured before), the solar quadrupole moment parameter $J_2$ to 1 part in 20, and the frame dragging effects on light due to the solar angular momentum to precision of 1 parts in $10^2$.

The LATOR mission technologically is a very sound concept; all technologies that are needed for its success have been already demonstrated as a part of the JPL's Space Interferometry Mission (SIM) development. (Accuracy of 5 picometers was already demonstrated in our SIM-related studies.) The LATOR concept arose from several developments at NASA and JPL that initially enabled optical astrometry and metrology, and also led to developing expertize needed for the precision gravity experiments. Technology that has become available in the last several years such as low cost microspacecraft, medium power highly efficient solid state lasers, and the development of long range interferometric techniques make possible an unprecedented factor of 30,000 improvement in this test of general relativity possible. This mission is unique and is the natural next step in solar system gravity experiments which fully exploits modern technologies.

LATOR will lead to very robust advances in the tests of Fundamental physics:  this mission could discover a violation or extension of general relativity, or reveal the presence of an additional long range interaction in the physical law.  With this mission testing theory to several orders of magnitude higher precision, finding a violation of general relativity or discovering a new long range interaction could be one of this era's primary steps forward in Fundamental physics. There are no analogs to the LATOR experiment; it is unique and is a natural culmination of solar system gravity experiments.  

This paper organized as follows: Section \ref{sec:motzz} provides more information about the theoretical framework, the PPN formalism, used to describe the gravitational experiments in the solar system. This section also summarizes the science motivation for the precision tests of gravity that recently became available.  Section \ref{sec:lator_description} provides the overview for the LATOR experiment including the preliminary mission design. In Section \ref{sec:conc} we discuss the next steps that will taken in the development of the LATOR mission. 

\section{SCIENTIFIC MOTIVATION}
\label{sec:motzz}
\subsection{PPN Parameters and Their Current Limits}

Generalizing on a phenomenological parameterization of the gravitational metric tensor field which Eddington originally developed for a special case, a method called the parameterized post-Newtonian (PPN) metric has been developed (see \cite{[41],[38],[39],[40],[59],[60]}).  This method  represents the gravity tensor's potentials for slowly moving bodies and weak interbody gravity, and valid for a broad class of metric theories including general relativity as a unique case.  The several parameters in the PPN metric expansion vary from theory to theory, and they are individually associated with various symmetries and invariance properties of underlying theory.  Gravity experiments can be analyzed in terms of the PPN metric, and an ensemble of experiments will determine the unique value for these parameters, and hence the metric field, itself.

In locally Lorentz-invariant theories the expansion of the metric field for a single, slowly-rotating gravitational source in PPN parameters is given by:
\begin{eqnarray}
g_{00}&=&1-2\frac{M}{r}\left(1-J_2\frac{R^2}{r^2}\frac{3\cos^2\theta-1}{2}\right)+
\nonumber \\[-4pt]
&&\hskip 8pt 
+~ 2\beta\frac{M^2}{r^2}+{\cal O}(c^{-6}),\nonumber\\[-4pt]
g_{0i}&=& 2(\gamma+1)\frac{[\vec{J}\times \vec{r}]_i}{r^3}+
{\cal O}(c^{-5}),\nonumber\\[-4pt]
g_{ij}&=&-\delta_{ij}\big[1+2\gamma \frac{M}{r} \left(1-J_2\frac{R^2}{r^2}\frac{3\cos^2\theta-1}{2}\right)+
\nonumber\\[-4pt]
&&\hskip 8pt + \frac{3}{2}\delta \frac{M^2}{r^2}\Big]+{\cal O}(c^{-6}),
\end{eqnarray}

\noindent where $M$ is the mass of the Sun, $R$ is the radius of the Sun, $\vec J$ is the angular momentum of the Sun, $J_2$ is the quadrupole moment of the Sun. $r$ is the distance between the observer and the center of the Sun.  $\beta, \gamma, \delta$ are the PPN parameters and in GR they are all equal to l. The term $M/r$ in the $g_{00}$ equation is the Newtonian limit; the terms multiplied by the post-Newtonian parameters $\beta, \gamma$,  are post-Newtonian terms. The term multiplied by the post-post-Newtonian parameter  $\delta$ also enters the calculation of the relativistic light deflection.

This PPN expansion serves as a useful framework to test relativistic gravitation in the context of the LATOR mission. In the special case, when only two PPN parameters ($\gamma$, $\beta$) are considered, these parameters have clear physical meaning. Parameter $\gamma$  represents the measure of the curvature of the space-time created by a unit rest mass; parameter  $\beta$ is a measure of the non-linearity of the law of superposition of the gravitational fields in the theory of gravity. GR, which corresponds to  $\gamma = \beta$  = 1, is thus embedded in a two-dimensional space of theories. The Brans-Dicke is the best known theory among the alternative theories of gravity.  It contains, besides the metric tensor, a scalar field and an arbitrary coupling constant $\omega$, which yields the two PPN parameter values $\gamma = (1+ \omega)/(2+ \omega)$, and $\beta$  = 1.  More general scalar tensor theories yield values of $\beta$ different from one.

PPN formalism proved to be a versatile method to plan gravitational experiments in the solar system and to analyze the data obtained \cite{[6],[9],[38],[39],[40],[41],[60],[44],ken_icarus95,[58],[59]}.  Different experiments test different combinations of these parameters (for more details, see \cite{[59]}). Until recently, the most precise value for the PPN parameter  $\gamma$ is at present given by Eubanks et al \cite{[23]} as: $|\gamma -1| = 0.0003$, which was obtained by means of astrometric VLBI. The secular trend of Mercury's perihelion, when describe in the PPN formalism, depends on another linear combination of the PPN parameters $\gamma$  and  $\beta$ and the quadrupole coefficient $J_{2\odot}$ of the solar gravity field:  $\lambda_\odot = (2 + 2\gamma -\beta)/3 + 0.296\times J_{2\odot}\times 10^4$. The combination of parameters $\lambda_\odot = 0.9996\pm 0.0006$, was obtained with the Mercury ranging data \cite{[31]}. The PPN formalism has also provided a useful framework for testing the violation of the Strong Equivalence Principle (SEP) for gravitationally bound bodies.  In that formalism, the ratio of passive gravitational mass $M_G$ to inertial mass $M_I$ of the same body is given by $M_G/M_I = 1 -\eta U_G/(M_0c^2)$, where $M_0$ is the rest mass of this body and $U_G$ is the gravitational self-energy. The SEP violation is quantified by the parameter $\eta$, which is expressed in terms of the basic set of PPN parameters by the relation $\eta = 4\beta-\gamma-3$. Analysis of planetary ranging data recently yielded an independent determination of parameter $\gamma$ \cite{[3],[2],[63]}: $|\gamma -1| = 0.0015 \pm 0.0021$; it also gave   with accuracy at the level of $|\beta -1| = -0.0010 \pm 0.0012$. With LLR finding that Earth and Moon fall toward the Sun at rates equal to 1.5 parts in 10$^{13}$, even in a conservative scenario where a composition dependence of acceleration rates masks a gravitational self energy dependence $\eta$ is constrained to be less than 0.0008 \cite{[2]}; without such accidental cancelation the $\eta$ constraint improves to 0.0003. Finally, the recent conjuction experiments with cassini spacecraft determined $\gamma$ with a truly remarkable accuracy of $\gamma-1=(2.1\pm2.3)\times10^{-5}$ \cite{cassini_ber}, thus opening a new exciting area of relativistic gravity tests.

The technology has advanced to the point that one can consider carrying out direct tests in a weak field to second order in the field strength parameter, $GM/Rc^2$. Although any measured anomalies in first or second order metric gravity potentials will not determine strong field gravity, they would signal that modifications in the strong field domain will exist.  The converse is perhaps more interesting:  if to high precision no anomalies are found in the lowest order metric potentials, and this is reinforced by finding no anomalies at the next order, then it follows that any anomalies in the strong gravity environment are correspondingly quenched. We shell discuss the recent motivations for the precision gravity tests below in more details.

\subsection{Motivations for Precision Gravity Experiments} 
\label{sec:mot}

After almost ninety years since general relativity was born, Einstein's theory has survived every test. Such a longevity, along with the absence of any adjustable parameters, does not mean that this theory is absolutely correct, but it serves to motivate more accurate tests to determine the level of accuracy at which it is violated. A significant number of these tests  were conducted over the period of last 35 years. As an upshot of these efforts, most alternative theories have been put aside; only those theories of gravity flexible enough have survived, the accommodation being provided by the free parameters and the coupling constant of the theory.  

Recently considerable interest has been shown in the physical processes occurring in the strong gravitational field regime. It should be noted that general relativity and some other alternative gravitational theories are in good agreement with the experimental data collected from the relativistic celestial mechanical extremes provided by the relativistic motions in the binary millisecond pulsars.  However, many modern theoretical models, which include general relativity as a standard gravity theory, are faced with the problem of the unavoidable appearance of space-time singularities. It is generally suspected that the classical description, provided by general relativity, breaks down in a domain where the curvature is large, and, hence, a proper understanding of such regions requires new physics. 

The continued inability to merge gravity with quantum mechanics indicate that the pure tensor gravity of general relativity needs modification or augmentation. The tensor-scalar theories of gravity, where the usual general relativity tensor field coexists with one or several long-range scalar fields, are believed to be the most promising extension of the theoretical foundation of modern gravitational theory. The superstring, many-dimensional Kaluza-Klein, and inflationary cosmology theories have revived interest in the so-called `dilaton fields', i.e. neutral scalar fields whose background values determine the strength of the coupling constants in the effective four-dimensional theory. The importance of such theories is that they provide a possible route to the quantization of gravity. Although the scalar fields naturally appear in the theory, their inclusion predicts different relativistic corrections to Newtonian motions in gravitating systems. These deviations from GR lead to a violation of the Equivalence Principle (either weak or strong or both), modification of large-scale gravitational phenomena, and generally lead to space and time variation of physical ``constants.'' As a result, this progress has provided new strong motivation for high precision relativistic gravity tests.

\begin{figure}[t!]
 \begin{center}
\noindent    
\epsfig{figure=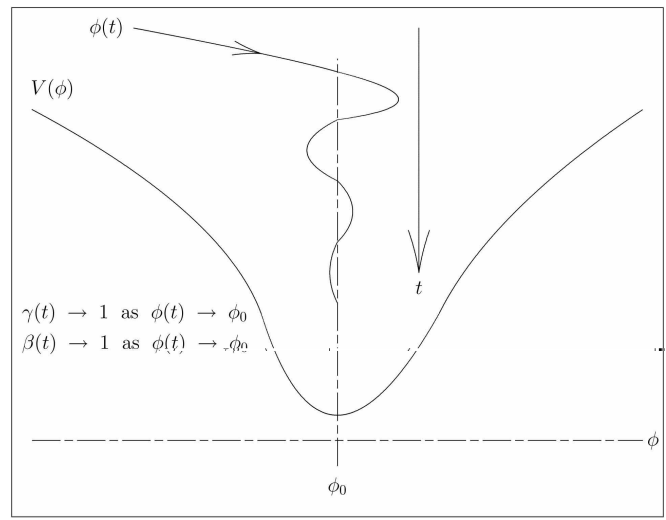,width=87mm}
\end{center}
\vskip -10pt 
  \caption{Typical cosmological dynamics of a background scalar field is shown if that field's coupling function to matter, $V(\phi)$, has an attracting point $\phi_0$. The strength of the scalar interaction's coupling to matter is proportional to the derivative (slope) of the coupling function, so it weakens as the attracting point is approached, and both the Eddington parameters $\gamma$ and $\beta$ (and all higher structure parameters as well)  approach their pure tensor gravity values in this limit.  But a small residual scalar gravity should remain today because this dynamical process is not complete, and that is what experiment seeks to find.
 \label{fig:attract}}
\end{figure}

The recent theoretical findings suggest that the present agreement between Einstein's theory and experiment might be naturally compatible with the existence of a scalar contribution to gravity. In particular, Damour and Nordtvedt \cite{[12a],[12]} (see also \cite{[10],[11a],[15],[16]} for non-metric versions of this mechanism) have recently found that a scalar-tensor theory of gravity may contain a `built-in' cosmological attractor mechanism towards GR.  A possible scenario for cosmological evolution of the scalar field was given in \cite{[12],[12n]}. Their speculation assumes that the parameter  $\frac{1}{2}(1-\gamma)$  was of order of 1 in the early universe, at the time of inflation, and has evolved to be close to, but not exactly equal to, zero at the present time (Figure \ref{fig:attract} illustrates this mechanism in more details). The expected deviation from zero may be of order of the inverse of the redshift of the time of inflation, or somewhere between 1 part per $10^5$ and 1 part per $10^7$ depending on the total mass density of the universe:  $1-\gamma \sim 7.3 \times 10^{-7}(H_0/\Omega_0^3)^{1/2}$, where $\Omega_0$ is the ratio of the current density to the closure density and $H_0$ is the Hubble constant in units of 100 km/sec/Mpc. This recent work in scalar-tensor extensions of gravity which are consistent with, indeed often part of, present cosmological models motivate new searches for very small deviations of relativistic gravity in the solar system, at levels of 10$^{-5}$ to 10$^{-7}$ of the post-Newtonian effects.   

The theoretical arguments above have been unexpectedly joined by a number of experimental results that motivate more precise gravitational experiments. Among those are the recent cosmological discoveries and the possible time variation detected in the fine structure constant. In particular, recent astrophysical measurements of the angular structure of the cosmic microwave background \cite{[20]}, the masses of large-scale structures \cite{[29]}, and the luminosity distances of type Ia supernovae \cite{[2c],[52]} have placed stringent constraints on the cosmological constant $\Lambda$ and also have led to a revolutionary conclusion: the expansion of the universe is accelerating. 
 The implication of these observations for cosmological models is that a classically evolving scalar field currently dominates the energy density of the universe. Such models have been shown to share the advantages of  $\Lambda$:  compatibility with the spatial flatness predicted inflation; a universe older than the standard Einstein-de Sitter model; and, combined with cold dark matter, predictions for large-scale structure formation in good agreement with data from galaxy surveys. Compared to the cosmological constant, these scalar field models are consistent with the supernovae observations for a lower matter density, $\Omega_0\sim 0.2$, and a higher age, $(H_0 t_0) \approx 1$. If this is indeed the case, the level $\gamma -1 \sim 10^{-6}-10^{-7}$ would be the lower bound for the present value of PPN parameter $\gamma$ \cite{[12a],[12]}. Combined with the fact that scalar field models imprint distinctive signature on the cosmic microwave background (CMB) anisotropy, they remain currently viable and should be testable in the near future. 

This completely unexpected discovery demonstrates the importance of testing the important ideas about the nature of gravity. We are presently in the ``discovery'' phase of this new physics, and while there are many theoretical conjectures as to the origin of a non-zero $\Lambda$, it is essential that we exploit every available opportunity to elucidate the physics that is at the root of the observed phenomena.

There is also experimental evidence for time-variability in the fine structure constant, $\alpha$, at the level of $\dot\alpha/(\alpha H_0) \sim 10^{-5}$  \cite{[34],[8]}. This is very similar to time variation in the gravitational constant, which is at the post-Newtonian level expressed as $\dot G/(G H_0)\approx \eta = 4\beta-\gamma-3$, thus providing a tantalizing motivation for further tests of the SEP (Strong Equivalence Principle) parameter $\eta$. A similar conclusion resulted from the recent analysis performed in \cite{[1],[7],[9]}. These new findings necessitate the measurements of $\gamma$  and $\beta$ in the range from $10^{-6}$ to $10^{-8}$ to test the corresponding gravitational scenario, thus requiring new gravitational physics missions. 

Even in the solar system, GR still faces challenges. There is the long-standing problem of the size of the solar quadrupole moment and its possible effect on the relativistic perihelion precession of Mercury (see review in \cite{[59]}). The interest in lies in the study of the behavior of the solar quadrupole moment versus the radius and the heliographic latitudes. This solar parameter has been very often neglected in the past, because it was rather difficult to determine an accurate value. The improvement of our knowledge of the accuracy of $J_2$ is certainly due to the fact that, today, we are able to take into account the differential rotation with depth. In fact, the quadrupole moment plays an important role in the accurate computation of several astrophysical quantities, such as the ephemeris of the planets or the general relativistic prediction for the precession of the perihelion of Mercury and other minor planets such as Icarus.  Finally, it is necessary to accurately know the value of the quadrupole moment to determinate the shape of the Sun, that is to say its oblateness. Solar oblateness measurements by Dicke and others in the past gave conflicting results for $J_2$ (reviewed on p. 145 of \cite{[16cw]}). A measurement of solar oblateness with the balloon-borne Solar Disk Sextant gave $J_2$ on the order of $2\times 10^{-7}$) \cite{lydon}. Helioseismic determinations using solar oscillation data have since implied a small value for $J_2$, on the order of $\sim 10^{-7}$, that is consistent with simple uniform rotation \cite{Brown,[27],[59]}. However, there exist uncertainties in the helioseismic determination for depths below roughly 0.4 $R_\odot$ which might permit a rapidly rotating core. LATOR can measure $J_2$ with accuracy sufficient to put this issue to rest.

Finally, there is now multiple evidence indicating that 70\% of the critical density of the universe is in the form of a ``negative-pressure'' dark energy component; there is no understanding as to its origin and nature. The fact that the expansion of the universe is currently undergoing a period of acceleration now seems inescapable: it is directly measured from the light-curves of several hundred type Ia supernovae \cite{[2c],[52],[3c]}, and independently inferred from observations of CMB by the WMAP satellite \cite{[4c]} and other CMB experiments \cite{[5c],[6c]}. Cosmic speed-up can be accommodated within general relativity by invoking a mysterious cosmic fluid with large negative pressure, dubbed dark energy. The simplest possibility for dark energy is a cosmological constant; unfortunately, the smallest estimates for its value are 55 orders of magnitude too large (for reviews see \cite{[7c],[8c]}). Most of the theoretical studies operate in the shadow of the cosmological constant problem, the most embarrassing hierarchy problem in physics. This fact has motivated a host of other possibilities, most of which assume $\Lambda=0$, with the dynamical dark energy being associated with a new scalar field (see \cite{[carroll]} and references therein). However, none of these suggestions is compelling and most have serious drawbacks. Given the challenge of this problem, a number of authors considered the possibility that cosmic acceleration is not due to some kind of stuff, but rather arises from new gravitational physics (see discussion in \cite{[21c],[17c],[24c],[carroll]}). In particular, extensions to general relativity in a low curvature regime were shown to predict an experimentally consistent universe evolution  without the need for dark energy. These dynamical models are expected to produce measurable contribution to the parameter $\gamma$  in experiments conducted in the solar system also at the level of $1-\gamma \sim 10^{-7}-10^{-9}$, thus further motivating the relativistic gravity research. Therefore, the PPN parameter $\gamma$ may be the only key parameter that holds the answer to most of the questions discussed. 

In summary, there are a number of theoretical reasons to question the validity of GR. Despite the success of modern gauge field theories in describing the electromagnetic, weak, and strong interactions, it is still not understood how gravity should be described at the quantum level. In theories that attempt to include gravity, new long-range forces can arise in addition to the Newtonian inverse-square law. Even at the purely classical level, and assuming the validity of the Equivalence Principle, Einstein's theory does not provide the most general way to generate the space-time metric. Regardless of whether the cosmological constant should be included, there are also important reasons to consider additional fields, especially scalar fields.   The LATOR mission is designed to address theses challenges.

\subsection{Look in the Near Future} 

Prediction of possible deviation of PPN parameters from the general relativistic values provides a robust theoretical paradigm and constructive guidance for experiments that would push beyond the present empirical upper bound on  $\gamma$  of $\gamma-1=(2.1\pm2.3)\times 10^{-5}$ (obtained by the Cassini conjunction experiments \cite{cassini_ber}). In addition to experiments, which probe parameter $\gamma$, any experiment pushing the present upper bounds on $\beta$  (i.e. $|\beta-1| < 5\times 10^{-4}$ from Anderson et al. \cite{[2],[63]} or LLR constraint on parameter $\eta = 4\beta-\gamma-3 \leq 3\times 10^{-4}$ \cite{[1],[6],[7],[2]}) will also be of great interest. Note that the Eddington parameter $\gamma$, whose value in general relativity is unity, is perhaps the most fundamental PPN parameter, in that $(1-\gamma)$ is a measure, for example, of the fractional strength of the scalar gravity interaction in scalar-tensor theories of gravity.  Within perturbation theory for such theories, all other PPN parameters to all relativistic orders collapse to their general relativistic values in proportion to $(1-\gamma)$. Therefore, measurement of the first order light deflection effect at the level of accuracy comparable with the second-order contribution would provide the crucial information separating alternative scalar-tensor theories of gravity from general relativity \cite{[40]}. 

By testing gravity at the level of accuracy needed to see the effects of the second order, one not simply discriminates among the alternative theories of gravity; in effect, one obtains the critical information on the beginning, current evolution and ultimate future of our universe.  The recent remarkable progress in observational cosmology has put general relativity at a test again by suggesting a non-Einsteinian model of universe's evolution. From the theoretical standpoint, the challenge is even stronger - if the gravitational field is to be quantized, the general theory of relativity will have to be modified. This is why the recent advances in the scalar-tensor extensions of gravity, that are consistent with the current inflationary model of the Big Bang, have motivated new search for a very small deviation of from Einstein's theory, at the level of three to five orders of magnitude below the level tested by experiment. 

\begin{figure*}[t!]
 \begin{center}
\noindent    
\psfig{figure=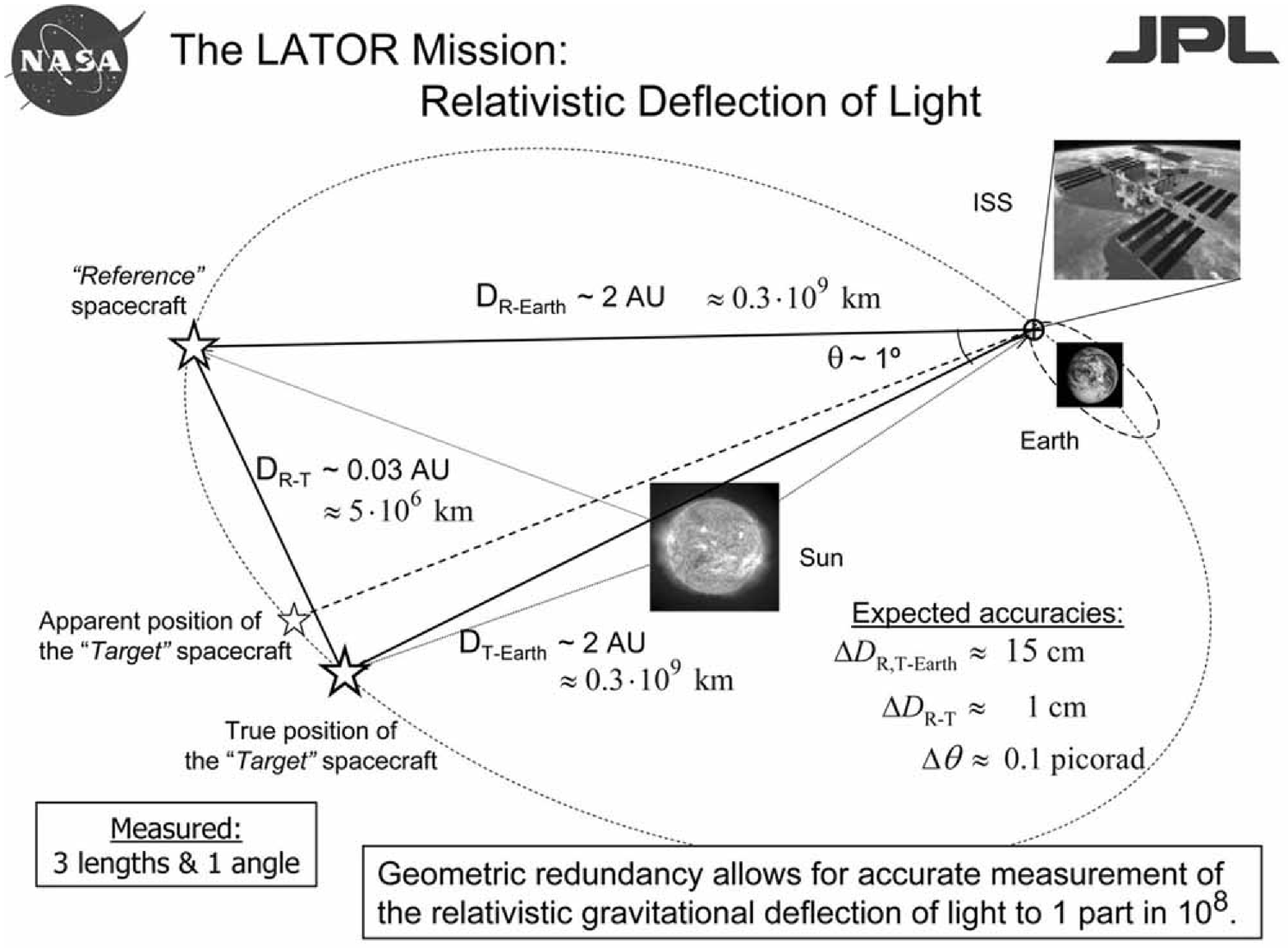,width=158mm}
\end{center}
\vskip -10pt 
  \caption{Geometry of the LATOR experiment to measure deviations from the Euclidean geometry in the solar gravity field. 
 \label{fig:lator}}
\end{figure*} 

 
Concluding, we point out that the recent progress in relativistic gravity research resulted in a significant tightening of the existing bounds on the PPN parameters obtained at the first post-Newtonian level of accuracy. However, this improvement is not sufficient to lead to groundbreaking tests of Fundamental physical laws addressed in Section \ref{sec:mot}. This is especially true, if the cosmological attractor discovered in \cite{[12],[12a]} is more robust, time variation in the fine structure constant will be confirmed in other experiments and various GR extensions will demonstrate feasibility of these methods for cosmology and relativistic gravity. The LATOR mission is proposed to directly address the challenges discussed above. We shall now discuss the LATOR mission in more details.

\section{OVERVIEW OF LATOR} 
\label{sec:lator_description}

The LATOR experiment uses laser interferometry between two micro-spacecraft (placed in heliocentric orbits, at distances $\sim$ 1 AU from the Sun) whose lines of sight pass close by the Sun to accurately measure deflection of light in the solar gravity. (A version of LATOR with a ground-based receiver was proposed in 1994 (performed under NRA 94-OSS-15) \cite{[53]}. Due to atmospheric turbulence and seismic vibrations that are not common mode to the receiver optics, a very long baseline interferometer (30 km) was proposed. This interferometer could only measure the differential light deflection to an accuracy of 0.1 $\mu$as, with a spacecraft separation of less than 1 arc minutes.) Another component of the experimental design is a long-baseline ($\sim100$ m) multi-channel stellar optical interferometer placed on the International Space Station (ISS). Figure \ref{fig:lator} shows the general concept for the LATOR missions including the mission-related geometry, experiment details  and required accuracies.

\begin{table*}[!ht!]
\caption{Comparable sizes of various light deflection effects in the solar gravity field.\label{tab:eff}}
\begin{center}
\begin{tabular}{|c|c|c|c|} \hline &&&\\[-8pt]
     Effect  & 
     Analytical Form      & 
     Value ($\mu$as)  &
     Value (pm) \\[3pt]
\hline \hline
&&&\\[-9pt]
   First  Order  &
   $2(1+\gamma )\frac{M}{R}$ & 
   $1.75\times 10^6$   & 
   $8.487\times10^{8}$      \\[4pt] 
&&&\\[-12pt] \hline 
&&&\\[-9pt]
   Second Order   &
   $([2(1+  \gamma)-\beta+\frac{3}{4} \delta]\pi- 
2(1+\gamma)^2)\frac{M^2}{R^2} $
   & 3.5   
   & 1702      \\[4pt] 
&&&\\[-12pt] \hline 
&&&\\[-8pt]
   Frame-Dragging   &
   $\pm 2(1+\gamma)\frac{J}{R^2}$ & 
   $\pm0.7$   &     $\pm339$ \\[4pt] 
&&&\\[-12pt] \hline 
&&&\\[-8pt]
   Solar Quadrupole  &
   $2(1+\gamma )J_2\frac{M}{R}$ & 
   0.2   &   97  \\[4pt] 
\hline 
\end{tabular}
\end{center}   

\end{table*}


\subsection{Mission Design} 

The LATOR mission consists of two low cost micro-spacecraft (the goal is to launch both spacecraft on a single Delta II launch vehicle). with three interferometric links between the craft and a beacon station on the ISS.  One of the longest arms of the triangle ($\sim$ 2 AU) passes near the Sun. The two spacecraft are in the helio-centric orbits and use lasers to measure the distance between them and a beacon station on the ISS. The laser light passes close to the Sun, which causes the light path to be both bent and lengthen. One spacecraft is at the limb of the Sun, the other one is $\sim 1^\circ$ away, as seen from the ISS. Each spacecraft uses laser ranging to measure the distance changes to the other spacecraft. The spatial interferometer is for measuring the angles between the two spacecraft and for the orbit determination purposes.

 As evident from Figure \ref{fig:lator}, the key element of the LATOR experiment is a redundant geometry optical truss to measure the departure from Euclidean geometry caused by Gravity.  The triangle in figure has three independent quantities but three arms are monitored with laser metrology. From three measurements one can calculate the Euclidean value for any angle in this triangle.  In Euclidean geometry these measurements should agree to high accuracy.  This geometric redundancy enables LATOR to measure the departure from Euclidean geometry caused by the solar gravity field to a very high accuracy. The difference in the measured angle and its Euclidean value is the non-Euclidean signal. To avoid having to make absolute measurements, the spacecraft are placed in an orbit where their impact parameters, the distance between the beam and the center of the Sun, vary significantly from $10 R_\odot$ to $1 R_\odot$ over a period of $\sim$ 20 days.

The shortening of the interferometric baseline (as compare to the previously studied version \cite{[53]}) is achieved solely by going into space to avoid the atmospheric turbulence and Earth's seismic vibrations. On the space station, all vibrations can be made common mode for both ends of the interferometer by coupling them by an external laser truss. This relaxes the constraint on the separation between the spacecraft, allowing it to be as large as few degrees, as seen from the ISS. Additionally, the orbital motion of the ISS provides variability in the interferometer's baseline projection as needed to resolve the fringe ambiguity of the stable laser light detection by an interferometer.

The first order effect of light deflection in the solar gravity caused by the solar mass monopole is 1.75 arcseconds (see Table \ref{tab:eff} for more details), which corresponds to a delay of $\sim$0.85 mm on a 100 m baseline. We currently are able to measure with laser interferometry distances with an accuracy (not just precision but accuracy) of $<$ 1 picometer. In principle, the 0.85 mm gravitational delay can be measured with $10^{-9}$ accuracy versus $10^{-4}$ available with current techniques. However, we use a conservative estimate for the delay of 10 pm which would produce the measurement of $\gamma$ to accuracy of 1 part in $10^{-8}$ (i.e improving the accuracy in determining this parameter by a factor of 30,000) rather than 1 part in $10^{-9}$. The second order light deflection is approximately 1700 pm and with 10 pm accuracy it could be measured with accuracy of $\sim1\times 10^{-3}$, including first ever measurement of the PPN parameter $\delta$.  The frame dragging effect would be measured with $\sim 1\times10^{-2}$ accuracy and the solar quadrupole moment (using the theoretical value of the solar quadrupole moment $J_2\simeq10^{-7}$) can be modestly measured to 1 part in 20, all with respectable signal to noise ratios.

The laser interferometers use $\sim$2W lasers and $\sim$20 cm optics for transmitting the light between spacecraft. Solid state lasers with single frequency operation are readily available and are relatively inexpensive.   For SNR purposes we assume the lasers are ideal monochromatic sources. For simplicity we assume the lengths being measured are 2AU = $3\times 10^8$ km. The beam spread is 1 $\mu$m/20 cm = 5 $\mu$rad (1 arcsecond). The beam at the receiver is $\sim$1,500 km in diameter, a 20 cm receiver will detect $1.71 \times 10^2$ photons/sec assuming 50\% q.e. detectors. 5 picometer (pm) resolution for a measurement of $\gamma$ to $\sim10^{-8}$ is possible with approximately 10 seconds of integration.

As a result, the LATOR experiment will be capable of measuring the angle between the two spacecraft to $\sim 0.01 ~\mu$as, which allows light deflection due to gravitational effects to be measured to one part in $10^8$. Measurements with this accuracy will lead to a better understanding of gravitational and relativistic physics. In particular, with LATOR, measurements of the first order gravitational deflection will be improved by a factor of 30,000. LATOR will also be capable of distinguishing between first order ($\sim M/R$) and second order ($\sim M^2/R^2$) effects. All effects, including the first and second order deflections, as well as the frame dragging component of gravitational deflection and the quadrupole deflection will be measured astrometrically.  We now outline the basic elements of the LATOR trajectory and optical design.

\subsection{Trajectory -- a 3:2 Earth Resonant Orbit} 

The objective of the LATOR mission includes placing two spacecraft into a heliocentric orbit with a one year period so that observations may be made when the spacecraft are behind the Sun as viewed from the ISS.  The observations involve the measurement of distance of the two spacecraft using an interferometer on-board the ISS to determine bending of light by the Sun.  The two spacecraft are to be separated by about 1$^\circ$, as viewed from the ISS.  

\begin{figure}[!ht!]
 \begin{center}
\noindent    
\psfig{figure=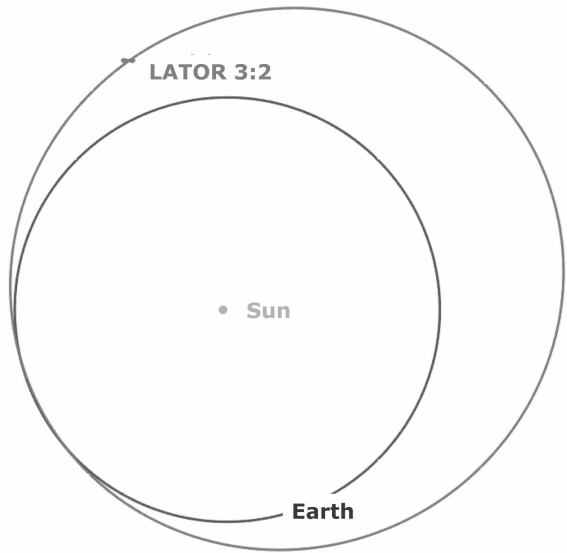,width=80mm}
\end{center}
\vskip -10pt 
  \caption{View from the North Ecliptic of the LATOR spacecraft in a 3:2 resonance. The epoch is taken near the first occultation.
 \label{fig:lator_traj1}}
\end{figure} 

One trajectory option would be to use a Venus flyby to place the spacecraft in a 1 yr orbit (perihelion at Venus orbit $\sim$0.73 AU and aphelion $\sim$1.27 AU). One complication of this approach is that the Venus orbit is inclined about 3.4$^\circ$ with respect to the ecliptic and the out-of-plane position of Venus at the time of the flyby determines the orbit inclination \cite{teamx}. The LATOR observations require that the spacecraft pass directly behind the Sun, i.e., with essentially no orbit inclination.  In order to minimize the orbit inclination, the Venus' flyby would need to occur near the time of Venus nodal crossing (i.e., around 7/6/2011). An approach with a type IV trajectory and a single Venus flyby requires a powered Venus flyby with about 500 to 900 m/s.  However, a type I trajectory to Venus with two Venus gravity assists would get LATOR into a desirable 1 year orbit at Earth's opposition.  This option requires no $\Delta v$ and provides repeated opportunities for the desired science observations. At the same time this orbit has a short launch period $\sim$17 days which motivated us to look for an alternative.

\begin{figure*}[t!]
 \begin{center}
\noindent 
\psfig{figure=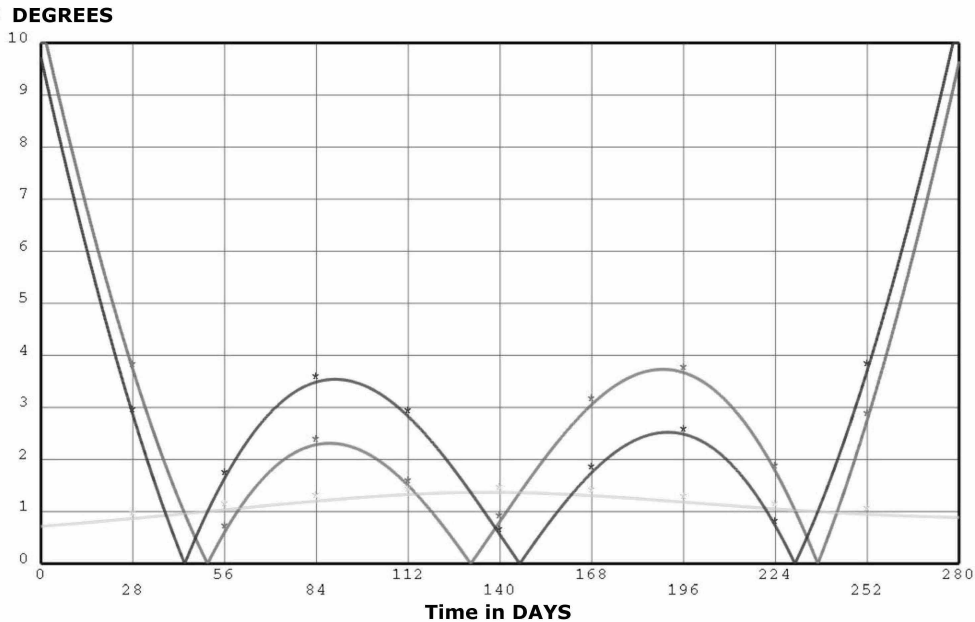,width=130mm}
\end{center}
\vskip -10pt 
  \caption{The Sun-Earth-Probe angle during the period of 3 occultations (two periodic curves) and the angular separation of the spacecraft as seen from the Earth (lower smooth line). Time shown is in days from the moment when one of the spacecraft is at 10$^\circ$ distance from the Sun. 
 \label{fig:lator_traj2}}
\end{figure*}

An good alternative to the double Venus flyby scenario was found when we studied  a possibility of launching LATOR into the orbit with a 3:2 resonance with the Earth \cite{teamx}. (The 3:2 resonance occurs when the Earth does 3 revolutions around the Sun while the spacecraft does exactly 2 revolutions of a 1.5 year orbit. The exact period of the orbit may vary slightly ($<$1\%) from a 3:2 resonance depending on the time of launch.) For this orbit, in 13 months after the launch, the spacecraft are within $\sim10^\circ$ of the Sun with first occultation occuring in 15 months after launch (See Figures \ref{fig:lator_traj1} and \ref{fig:lator_traj2}).  At this point, LATOR is orbiting at a slower speed than the Earth, but as LATOR approaches its perihelion, its motion in the sky begins to reverse and the spacecraft is again occulted by the Sun 18 months after launch.  As the spacecraft slows down and moves out toward aphelion, its motion in the sky reverses again and it is occulted by the Sun for the third and final time 21 months after launch.  This entire process will again repeat itself in about 3 years after the initial occultation, however, there may be a small maneuver required to allow for more occultations.  Therefore, to allow for more  occultations in the future, there may be a need for an extra few tens of m/s of $\Delta v$. 

The  $C_3$ required for launch will vary between $\sim$10.6 km$^2$/s$^2$ -- 11.4 km$^2$/s$^2$ depending on the time of launch, but it is suitable for a Delta II launch vehicle.   The desirable $\sim1^\circ$ spacecraft separation  (as seen from the Earth) is achieved by performing a 30 m/s maneuver after the launch.  This results in the second spacecraft being within $\sim$ 0.6$^\circ$ -- 0.9$^\circ$ separation during the entire period of 3 occultations by the Sun.

Figures \ref{fig:lator_traj1} and \ref{fig:lator_traj2} show the  trajectory  and the occultations in more details.  The first figure is the spacecraft position in the solar system showing the Earth's and LATOR's orbits (in the 3:2 resonance) relative to the Sun.  The epoch of this figure shows the spacecraft passing behind the Sun as viewed from the Earth.  The second figure shows the trajectory when the spacecraft would be within 10 degrees of the Sun as viewed from the Earth.  This period of 280 days will occur once every 3 years, provided the proper maneuvers are performed.  The two similar periodic curves give the Sun-Earth-Probe angles for the 2 spacecraft while the lower smooth curve gives the angular separation of the spacecraft as seen from the Earth. We intend to further study this trajectory as the baseline option for the LATOR mission.

\subsection{ Optical Design} 

A single aperture of the interferometer on the ISS consists of three 10 cm diameter telescopes. One of the telescopes with a very narrow bandwidth laser line filter in front and with an InGAs camera at its focal plane, sensitive to the 1.3 $\mu$m laser light, serves as the acquisition telescope to locate the spacecraft near the Sun.

The second telescope emits the directing beacon to the spacecraft. Both spacecraft are served out of one telescope by a pair of piezo controlled mirrors placed on the focal plane. The properly collimated laser light ($\sim$10W) is injected into the telescope focal plane and deflected in the right direction by the piezo-actuated mirrors. 

The third telescope is the laser light tracking interferometer input aperture which can track both spacecraft at the same time. To eliminate beam walk on the critical elements of this telescope, two piezo-electric X-Y-Z stages are used to move two single-mode fiber tips on a spherical surface while maintaining focus and beam position on the fibers and other optics. Dithering at a few Hz is used to make the alignment to the fibers and the subsequent tracking of the two spacecraft completely automatic. The interferometric tracking telescopes are coupled together by a network of single-mode fibers whose relative length changes are measured internally by a heterodyne metrology system to an accuracy of less than 10 picometer.

The spacecraft  are identical in construction and contain a relatively high powered (2 W), stable (2 MHz per hour $\sim$  500 Hz per second), small cavity fiber-amplified laser at 1.3 $\mu$m. Three quarters of the power of this laser is pointed to the Earth through a 20 cm aperture telescope and its phase is tracked by the interferometer. With the available power and the beam divergence, there are enough photons to track the slowly drifting phase of the laser light. The remaining part of the laser power is diverted to another telescope, which points towards the other spacecraft. In addition to the two transmitting telescopes, each spacecraft has two receiving telescopes.  The receiving telescope on the ISS, which points towards the area near the Sun, has laser line filters and a simple knife-edge coronagraph to suppress the Sun light to 1 part in 10,000 of the light level of the light received from the space station. The receiving telescope that points to the other spacecraft is free of the Sun light filter and the coronagraph.

In addition to the four telescopes they carry, the spacecraft also carry a tiny (2.5 cm) telescope with a CCD camera. This telescope is used to initially point the spacecraft directly towards the Sun so that their signal may be seen at the space station. One more of these small telescopes may also be installed at right angles to the first one to determine the spacecraft attitude using known, bright stars. The receiving telescope looking towards the other spacecraft may be used for this purpose part of the time, reducing hardware complexity. Star trackers with this construction have been demonstrated many years ago and they are readily available. A small RF transponder with an omni-directional antenna is also included in the instrument package to track the spacecraft while they are on their way to assume the orbital position needed for the experiment. 

The LATOR experiment has a number of advantages over techniques which use radio waves to measure gravitational light deflection. Advances in optical communications technology, allow low bandwidth telecommunications with the LATOR spacecraft without having to deploy high gain radio antennae needed to communicate through the solar corona. The use of the monochromatic light enables the observation of the spacecraft almost at the limb of the Sun, as seen from the ISS. The use of narrowband filters, coronagraph optics and heterodyne detection will suppress background light to a level where the solar background is no longer the dominant noise source. In addition, the short wavelength allows much more efficient links with smaller apertures, thereby eliminating the need for a deployable antenna. Finally, the use of the ISS will allow conducting the test above the Earth's atmosphere -- the major source of astrometric noise for any ground based interferometer. This fact justifies LATOR as a space mission.
 
\section{ CONCLUSIONS} 
\label{sec:conc}
 
The LATOR mission aims to carry out a test of the curvature of the solar system's gravity  field with an accuracy better than 1 part in 10$^{8}$. In spite of the previous space missions exploiting radio waves for tracking the spacecraft, this mission manifests an actual breakthrough in the relativistic gravity experiments as it allows to take full advantage of the optical techniques that recently became available.  Our next steps will be to perform studies of trajectory configuration and  conduct a mission design including the launch vehicle choice trade studies. Our analysis will concentrate on the thermal design of the instrument; analysis of the launch options and configuration; estimates of on-board power and weight requirements; as well as analysis of optics and vibration contamination for the interferometer.  We also plan to  develop an end-to-end mission simulation, including detailed astrometric model and the mission error budget. 

The LATOR experiment technologically is a very sound concept; all technologies that are needed for its success have been already demonstrated as a part of the JPL's Space Interferometry Mission development.  The concept arose from several developments at NASA and JPL that initially enabled optical astrometry and metrology, and also led to developing expertize needed for the precision gravity experiments. Technology that has become available in the last several years such as low cost microspacecraft, medium power highly efficient solid state lasers for space applications, and the development of long range interferometric techniques make the LATOR mission feasible. The LATOR experiment does not need a drag-free system, but uses a geometric redundant optical truss to achieve a very precise determination of the interplanetary distances between the two micro-spacecraft and a beacon station on the ISS. The interest of the approach is to take advantage of the existing space-qualified optical technologies leading to an outstanding performance in a reasonable mission development time.  The  availability of the space station makes this mission concept realizable in the very near future; the current mission concept calls for a launch as early as in 2009 at a cost of a NASA MIDEX mission.   

LATOR will lead to very robust advances in the tests of Fundamental physics: this mission could discover a violation or extension of general relativity, or reveal the presence of an additional long range interaction in the physical law.  There are no analogs to the LATOR experiment; it is unique and is a natural culmination of solar system gravity experiments. 

\section*{ Acknowledgments} 

The authors would like to thank Yekta Gursel of JPL for many fruitful discussions. The work described here was carried out at the Jet Propulsion Laboratory, California Institute of Technology, under a contract with the National Aeronautic and Space Administration.



\end{document}